\documentclass[journal] {IEEEtran}

\usepackage{epsfig}
\usepackage{color}
\usepackage{cite}

\begin{document}

\title{On Impedance Bandwidth of Resonant Patch Antennas Implemented Using
Structures with Engineered Dispersion}

\author{Pekka~M.~T.~Ikonen,~\IEEEmembership{Student Member,~IEEE}, Pekka Alitalo,
        Sergei~A.~Tretyakov,~\IEEEmembership{Senior Member,~IEEE}
\thanks{The authors are with the Radio Laboratory/SMARAD Centre of Excellence,
Helsinki University of Technology, P.O. Box 3000, FI-02015 TKK, Finland. (e-mail: pekka.ikonen@tkk.fi).}}

\maketitle
\begin{abstract}
We consider resonant patch antennas, implemented using loaded transmission-line networks and other exotic
structures having engineered dispersion. An analytical expression is derived for the ratio of radiation quality
factors of such antennas and conventional patch antennas loaded with (reference) dielectrics. In the ideal case
this ratio depends only on the propagation constant and wave impedance of the structure under test, and it can
be conveniently used to study what kind of dispersion leads to improved impedance bandwidth. We illustrate the
effect of dispersion by implementing a resonant patch antenna using a periodic network of LC elements. The
analytical results predicting enhanced impedance bandwidth compared to the reference results are validated using
a commercial circuit simulator. Discussion is conducted on the practical limitations for the use of the proposed
expression.
\end{abstract}


\section{Introduction}

Size reduction of patch antennas using different materials filling the volume under the antenna element is a
classical miniaturization technique \cite{Bahl}. Typically, conventional dielectric substrates have been used to
decrease the physical dimensions of the radiator, and it is known that the impedance bandwidth decreases roughly
inversely with increasing the effective permittivity of the substrate (with a fixed substrate height)
\cite{Jackson}. It has been observed, however, that increase of the effective substrate permeability has only a
small effect on the bandwidth, if the permeability is assumed to be lossless and dispersion-free \cite{Jackson,
Hansen}. This observation has lead to a flow of research concerning patch antenna miniaturization using
magnetodielectric substrates, e.g.~\cite{Mosallaei_disagree}--\cite{Pekka2}. Recently, also several other exotic
materials have been proposed for size reduction: For example, authors of \cite{Mosallaei} proposed to use a
reactive impedance substrate, and authors of \cite{mah, Tretyakov} studied miniaturization using double-negative
metamaterials.

From the commonly used transmission-line (TL) representation of a patch antenna \cite{Bahl} it is clear that the
purpose of different loading materials is to optimize the input susceptance seen at the antenna terminal, while
maintaining the radiation conductance at a reasonable level. If the antenna geometry leaves the radiation
conductance intact, the dispersion characteristics of the loading material solely determine the impedance
bandwidth performance of the antenna under test (AUT) compared to the performance of the same-size reference
antenna, miniaturized using normal dielectrics. Authors of \cite{Pekka1, Pekka2} derived an analytical
expression for the ratio of radiation quality factors ($Q^{\rm rel}_{\rm r}$) of the above referred antennas
(resonant AUT and the reference antenna), and used this ratio as a figure of merit to test the suitability of
different materials in efficient antenna miniaturization. $Q^{\rm rel}_{\rm r}$ derived in \cite{Pekka1,Pekka2}
is solely a function of the effective AUT (substrate) material parameters, and it can be used when the material
is homogeneous and it is meaningful to assign effective material parameters (for other criteria, please see
\cite{Pekka2}).

Recently, there have been suggestions to manipulate the dispersion characteristics of patch antennas by
implementing the actual antennas as periodic networks of LC elements \cite{Schussler}--\cite{Lee}. The aim has
been to reduce the resonant length by synthesizing large propagation constants for the network. One interesting
related idea would be to synthesize a network whose wave impedance and propagation constant resemble those of
weakly dispersive high-permeability materials. If the implemented LC network comprising the antenna would still
allow effective radiation (the radiation conductance would not deteriorate due to implementation details), one
could possibly benefit from the advantages of magnetic loading \cite{Hansen} and weaken the strong ``negative''
effect of frequency dispersion \cite{Pekka1}. This expectation follows from the important fact that the
performance of the network is not based on resonant inclusions.

To be able to synthesize the desired dispersion characteristics for the network it would be convenient to have a
figure of merit that would immediately tell if the proposed dispersion allows the realization of a wider
impedance bandwidth than reference dielectrics. If the unit cell dimension is not very small, it might be more
meaningful to characterize the network using its effective propagation constant and wave impedance, rather than
the extracted material parameters \cite{Caloz}. Thus, $Q_{\rm r}^{\rm rel}$ proposed in \cite{Pekka1, Pekka2}
might not be directly applicable. Here we derive a figure of merit, similar to that in \cite{Pekka1, Pekka2},
but expressed as a function of the effective wave impedance and propagation constant of the structure/material
under test. This figure of merit is used to design a patch antenna with enhanced impedance bandwidth compared to
the reference antenna using a periodic network of LC elements.

\section{Formulation}

\begin{figure}[t!]
\centering \epsfig{file=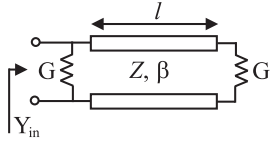, width=3.5cm} \caption{Patch antenna represented as a transmission-line section.}
\label{TL}
\end{figure}

Consider a rectangular patch antenna having width $w$, height $h$, and length $l$. The volume under the patch
element is loaded with a certain \emph{low-loss} material. The TL-representation of the antenna is shown in
Fig.~\ref{TL}. Using standard TL-equations we can write the input admittance of the antenna in the following
form:
\begin{equation}
Y_{\rm in} =  G  + \frac{G[1 + j\frac{Y}{G}\tan(\beta l)]}{1 + j\frac{G}{Y}\tan(\beta l)}, \label{Yin1}
\end{equation}
where $Y = 1/Z$ is the characteristic admittance of the equivalent TL, $G$ is the radiation conductance, and
$\beta$ is the equivalent propagation constant of the antenna segment. When the antenna operates close to its
fundamental (parallel) resonance ($\beta l \approx \pi$), $\tan (\beta l)$ has a very small value. Thus, after
applying Taylor expansion to the denominator of (\ref{Yin1}) we can write the input admittance in the following
approximate form:
\begin{equation}
Y_{\rm in} \approx 2G + j\frac{Y^2 - G^2}{Y}\tan(\beta l).
\end{equation}
Typically the height of a resonant patch is a small fraction of the wavelength and noticeably smaller than the
width of the antenna. By applying standard expressions, typically used to estimate the characteristic admittance
and radiation conductance \cite{Bahl}, we observe that $Y^2\gg G^2$, and the input admittance can further be
simplified as
\begin{equation}
Y_{\rm in} \approx 2G + jY\tan(\beta l). \label{Yin}
\end{equation}
One of the definitions for the quality factor of a resonator operating close to its parallel resonance reads
\cite{Collin}
\begin{equation}
Q = \frac{\omega}{2G}\frac{\partial B}{\partial\omega}\bigg{|}_{\omega=\omega_x}, \label{Q}
\end{equation}
where $B$ is the input susceptance and $\omega_x$ is the operational angular frequency of the resonator. In our
case eq.~(\ref{Q}) is the radiation quality factor since only radiation losses are assumed. We get the following
expression for the quality factor after differentiating the input susceptance:
\begin{equation}
Q = \frac{\omega}{2G}\frac{l}{Z}\frac{\partial\beta}{\partial\omega}\bigg{|}_{\omega=\omega_x}. \label{Qc}
\end{equation}
Take a resonant patch antenna loaded with a dispersion-free dielectric material offering the same size reduction
[$k_0\sqrt{\epsilon_{\rm r}^{\rm ref}} = \beta(\omega_{\rm x})$] as a reference antenna. The ratio of the
quality factors of the two antennas becomes
\begin{equation}
Q^{\rm rel} = \frac{Q}{Q^{\rm ref}} = m  \frac{Z^{\rm ref}}{Z}\frac{\partial
\beta}{\partial\omega}\bigg{(}\frac{\partial \beta^{\rm ref}}{\partial\omega}\bigg{)}^{-1}, \label{Qr1}
\end{equation}
where we $m = G^{\rm ref}/G$ is the ratio of radiation conductances. In the quasi-static regime the propagation
constant and characteristic impedance for the reference case read \cite{Collin}
\begin{equation}
\beta^{\rm ref} = k_0\sqrt{\epsilon^{\rm ref}_{\rm r}}, \quad Z^{\rm ref} =
\eta_0\frac{h}{w}\frac{1}{\sqrt{\epsilon^{\rm ref}_{\rm r}}}, \label{bZ}
\end{equation}
where $k_0,\eta_0$ are the free space wave number and wave impedance, respectively. From the requirement that
both antennas should have the resonant length at the same frequency, we get a relation between $\beta$ and
$\epsilon^{\rm ref}$:
\begin{equation}
\sqrt{\epsilon_{\rm r}^{\rm ref}} = \frac{\beta}{\omega}c_0 = \frac{c_0}{v_{\rm ph}}, \label{eps}
\end{equation}
where $v_{\rm ph}$ is the phase velocity in the equivalent transmission-line of the AUT. Moreover, the relation
between the effective wave impedance $Z_{\rm w}$ of the AUT and the characteristic impedance reads in our
geometry \cite{Collin}
\begin{equation}
Z = \frac{h}{w}Z_{\rm w}. \label{Z}
\end{equation}
After this the ratio of quality factors takes the form
\begin{equation}
Q^{\rm rel} = \frac{Q}{Q^{\rm ref}} = m \frac{\eta_0}{c_0}\frac{v_{\rm ph}^2}{Z_{\rm w}}\frac{\partial
\beta}{\partial \omega}\bigg{|}_{\omega = \omega_{\rm x}}. \label{mod11}
\end{equation}

\subsubsection{Comparing the obtained expression with previously introduced expression}

The ratio of (radiation) quality factors of ideally shaped resonant patch antennas (AUT and reference antenna)
was derived in the following form in \cite{Pekka1,Pekka2}:
\begin{equation}
Q^{\rm rel} = \frac{Q}{Q^{\rm ref}} = \frac{1}{2\mu_{\rm r}}\bigg{(}\frac{1}{\mu_{\rm r}}\frac{\partial
(\omega\mu_{\rm r})}{\partial\omega} + \frac{1}{\epsilon_{\rm r}}\frac{\partial (\omega\epsilon_{\rm
r})}{\partial\omega}\bigg{)}\bigg{|}_{\omega = \omega_x}, \label{Pekka}
\end{equation}
where $\epsilon_{\rm r},\mu_{\rm r}$ are the AUT effective (substrate) material parameters (low losses are
assumed). To derive eq.~(\ref{Pekka}) authors of \cite{Pekka1, Pekka2} integrated the field energy density over
the volume of the antenna, and further applied standard definitions to calculate the quality factors using
stored electromagnetic energy and radiated power.

Let us in the following assume that $m=1$ and we can assign effective material parameters for the tested
material. We adopt the notation considered in \cite{Pekka1, Pekka2}:
\begin{equation}
\beta = \frac{\omega}{c_0}\sqrt{\mu_{\rm r}\epsilon_{\rm r}}, \quad Z = \eta_0\frac{h}{w}\sqrt{\frac{\mu_{\rm
r}}{\epsilon_{\rm r}}}.
\end{equation}
With the above notations eq.~(\ref{mod11}) transforms into
\begin{equation}
Q^{\rm rel} = \frac{Q}{Q^{\rm ref}} = \frac{c_0}{\epsilon_{\rm r}\mu_{\rm r}}\sqrt{\frac{\epsilon_{\rm
r}}{\mu_{\rm r}}}\frac{\partial \beta}{\partial\omega}\bigg{|}_{\omega=\omega_x}, \label{mod2}
\end{equation}
and after differentiating the propagation constant we get the same expression as (\ref{Pekka}).

\subsubsection{Comments related to the practical use of eq.~(\ref{mod11})}

When deriving eq.~(\ref{Pekka}) authors of \cite{Pekka1, Pekka2} assumed that $m=1$ since the considered
antennas had the same-size continuous metal plates acting as antenna elements, and the classical assumption
according to which the width of the radiating edge predominantly determines the radiation conductance
\cite{Bahl}, was used. However, in the real antenna implementations based on LC element networks the actual
(realizable) radiation properties might not be so easily predicted by standard TL-expressions: Authors of
\cite{Schussler} observed that the power delivered to the furthermost radiating slot was 2 dB lower than the
power delivered to the first slot. In addition to this, parasitic radiation from the inductors (affecting the
fringing field) was reported. Later authors of \cite{Schussler2} presented an empirical formula to estimate the
radiation conductance of antennas based on LC-networks, and an expression to estimate the maximum achievable
bandwidth. Authors of \cite{Lee} reported rather low radiation efficiencies, and speculated that the low
efficiencies were due to large currents concentrated near vias implemented as lossy conductors. In this case the
standing wave pattern on the antenna element might not be anymore a pure $\sin$-function as in a normal patch.
This has a natural implication on the radiated power since most of the radiation should still happen due to
fringing fields (represented as a conductance sitting at the patch edge where maximum of voltage should occur).

Due to the above uncertainties that depend highly on the particular implementation details, and the empirical
expression for $G$ found in \cite{Schussler2}, we assume in the following generally that $m\geq1$. With this
assumption we are able to study the dispersion characteristics of the tested structures/materials separately
from the radiation properties. If the proposed dispersion is desirable, one should still before the final
implementation make sure that the antenna radiates properly.

\section{Example of application: Patch antenna implemented as an LC network}

\begin{figure}[b!] \centering \epsfig{file=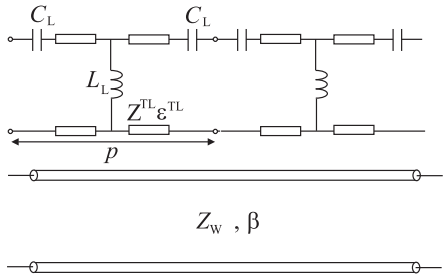, width=5.8cm} \caption{Transmission-line periodically
loaded with bulk capacitors and inductors. Equivalent homogeneous transmission-line having certain effective
wave impedance and propagation constant.} \label{circ}
\end{figure}

In this section we synthesize the performance of a compact patch antenna using a periodical arrangement of
loaded transmission-lines, Fig.~\ref{circ}. We use eq.~(\ref{mod11}) to choose practically realizable network
parameters so that $Q_{\rm r}^{\rm rel}$ goes below unity, i.e.~the TL-antenna is more wideband than the
same-size reference antenna. It is assumed that $m=1$ since we cannot at this stage readily predict the
influence of the practical realization to the equivalent radiation conductance or voltage wave pattern.

The Bloch impedance $Z_{\rm B}$ and propagation constant $\beta$ in the one-dimensional (1D) line are calculated
using the design equations available in \cite{PA}. Similarly as in \cite{Grbic} we relate the Bloch impedance to
the effective wave impedance in the equivalent homogeneous line:
\begin{equation}
Z_{\rm w} = \frac{p}{h}Z_{\rm B}, \label{Zw1}
\end{equation}
where $p$ is the unit cell period and $h$ is the height of the (network) substrate. With the help of
eq.~(\ref{mod11}) we choose the following parameters (notations are clear from Fig.~\ref{circ}): $p=5$ mm, $h=2$
mm, $Z^{\rm TL} = 200$ $\Omega$, $\epsilon_{\rm r}^{\rm TL}=2.33$, $C_{\rm L}=200$ pF, $L_{\rm L}=100$
nH.

\begin{figure}[t!]
\centering \epsfig{file=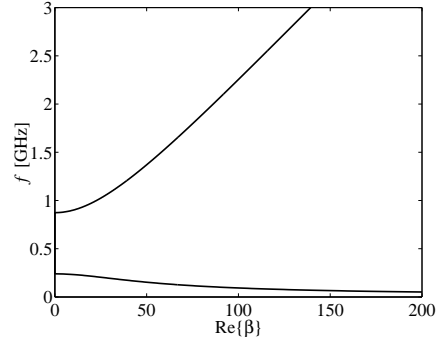, width=5.7cm} \caption{Frequency as function of the propagation constant in the
homogenized 1D transmission-line.} \label{b}
\end{figure}

\begin{figure}[t!]
\centering \epsfig{file=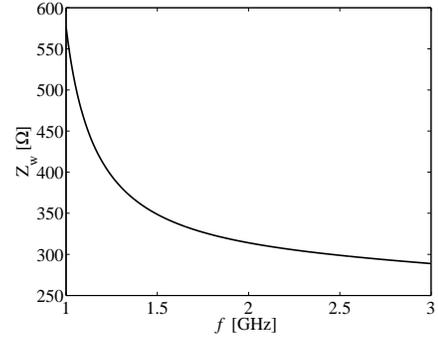, width=5.7cm} \caption{Wave impedance as a function of frequency in the
homogenized 1D transmission-line.} \label{Zww}
\end{figure}

\begin{figure}[b!]
\centering \epsfig{file=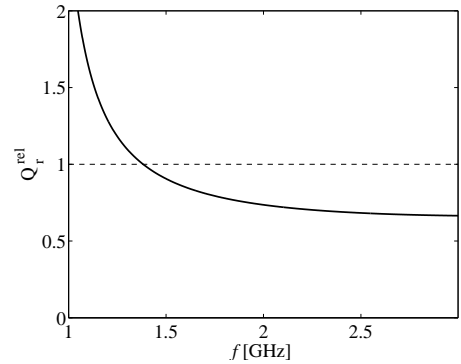, width=6cm} \caption{The relative radiation quality factor.} \label{Qrr}
\end{figure}

The equivalent propagation constant and wave impedance in the homogenized 1D line are shown in
Figs.~\ref{b},\ref{Zww}, respectively. The relative radiation quality factor [eq.~(\ref{mod11})] is shown in
Fig.~\ref{Qrr}. We can see that $Q_{\rm r}^{\rm rel}$ is below unity over a wide frequency band when the antenna
operates in the right-hand branch (forward wave) of the dispersion curve. Initial simulations with different
TL-structures\footnote{For which it is meaningful to assign effective material parameters due to very small
period $p$ compared to the wavelength.} have shown that the effective permeability is high and weakly dispersive
only in the right-hand branch. This is reflected also in the result depicted in Fig.~\ref{Qrr}, which resembles
the result obtained using weakly dispersive magnetic materials under the antenna element \cite{Pekka1, Pekka2}.

\begin{figure}[t!]
\centering \epsfig{file=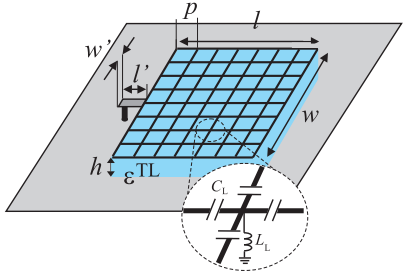, width=5.3cm} \caption{Schematic illustration of the patch antenna.} \label{sche}
\end{figure}

The designed antenna has length $l=7\times p=35$ mm and width $w=8\times p=40$ mm (see the schematic
illustration in Fig.~\ref{sche}) with the predicted resonant frequency around 2.5~GHz. The inductors and
capacitors have been removed from the last periods near the radiating edges to minimize their effect on the
fringing fields. We use data in Figs.~\ref{b}, \ref{Zww} to calculate the reflection coefficient for the
TL-antenna using the TL-model introduced in \cite{Pekka1, Pekka2}. The calculation is repeated for a reference
antenna loaded with dispersion-free dielectrics. In the TL-model, a short section of empty TL is left near the
radiating edges to reflect the absence of bulk components (the same empty space is left in the reference
calculation). In the calculations $l'=p$, and $w'$ has been varied to tune the quality of matching to be the
same with both antennas.

\begin{figure}[b!]
\centering \epsfig{file=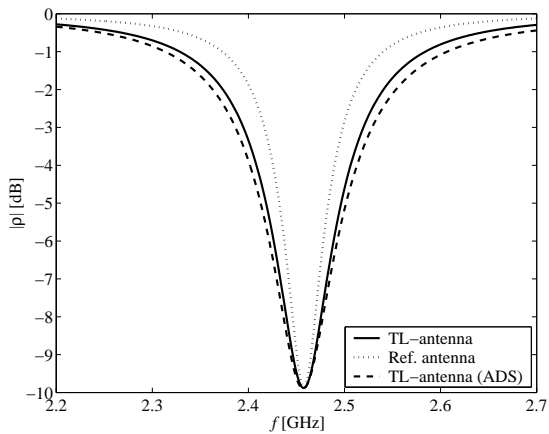, width=7.3cm} \caption{Reflection coefficient data for the transmission-line
antenna and the same size reference antenna.} \label{S11r}
\end{figure}

The results of calculations are depicted in Fig.~\ref{S11r}. We have also implemented the TL-antenna in the
Agilent ADS circuit simulator (no reference antenna simulations have been performed since the quantitative
accuracy of the TL-model for this case has been demonstrated earlier \cite{Pekka2}). Radiation has been modeled
both in the TL-model and in ADS by an equivalent radiation conductance $G=G^{\rm ref}=1/90(w/\lambda_0)^2$
\cite{Bahl}, see Fig.~\ref{TL}.
$Q_{\rm r}^{\rm rel}$ calculated from the reflection data is 0.66, whereas eq.~(\ref{mod11}) gives 0.69. The
result predicted by ADS closely agrees with the TL-result (with a fixed $G$) and validates the feasibility of
the proposed network (dispersion) in antenna miniaturization. We have to bear in mind, however, that in the
practical LC network implementation the radiation might be weaker \cite{Schussler, Lee} than predicted by the
standard expressions for $G$ given in \cite{Bahl}. This effect is, however, highly dependent on the
implementation details.

\section{Conclusion}

We have derived an explicit expression for the ratio of radiation quality factors of resonant patch antennas, 1)
implemented using structures having engineered dispersion and 2) loaded with conventional dielectrics. This
expression allows to study conveniently what kind of dispersion leads to efficient size reduction. As an
example, a patch antenna with desirable dispersion characteristics has been implemented using a periodical
network of LC elements.

\section*{Acknowledgement}
The authors wish to thank Dr.~S.~Maslovski for useful related discussions conducted in the spring 2005.
Mr.~J.~Holopainen, Mr.~T.~Kiuru, and Dr.~V.~M\"ott\"onen are acknowledged for additional simulations and help
with the simulation software.

\end{document}